\documentclass[reprint,aps,prl,showpacs,floatfix,notitlepage,superscriptaddress]{revtex4-2}
\usepackage{epsfig,graphics,amssymb,amsmath,subeqnarray,color,bm,bbm}
\usepackage{graphicx}
\usepackage[toc,page]{appendix}
\usepackage{float}
\usepackage[colorlinks=true,allcolors=blue]{hyperref}

\usepackage{gensymb} 
\let\epsilon\varepsilon
\usepackage{dcolumn}
\begin{document}

\title{Entropy production and thermodynamic inference for stochastic microswimmers} 

\author{Michalis Chatzittofi}
\address{Max Planck Institute for Dynamics and Self-Organization (MPI-DS), D-37077 Göttingen, Germany} 
\author{Jaime Agudo-Canalejo}
\email{jaime.agudo@ds.mpg.de}
\address{Max Planck Institute for Dynamics and Self-Organization (MPI-DS), D-37077 Göttingen, Germany} 
\author{Ramin Golestanian}
\email{ramin.golestanian@ds.mpg.de}
\address{Max Planck Institute for Dynamics and Self-Organization (MPI-DS), D-37077 Göttingen, Germany} 
\address{Rudolf Peierls Centre for Theoretical Physics, University of Oxford, Oxford OX1 3PU, United Kingdom}

\date{\today}

\begin{abstract}
The question of characterization of the degree of non-equilibrium activity in active matter systems is studied in the context of a stochastic microswimmer model driven by a chemical cycle. The resulting dynamical properties and entropy production rate unravel a complex interplay between the chemical and the hydrodynamic degrees of freedom beyond linear response, which is not captured by conventional phenomenological approaches. By studying the precision-dissipation trade-off, a new protocol is proposed in which microscopic chemical driving forces can be inferred  experimentally. Our findings highlight subtleties associated with the stochastic thermodynamics of autonomous microswimmers.
\end{abstract} 

\maketitle

Understanding entropy production and thermodynamic inference \cite{seifert2019stochastic} in autonomous systems \cite{Gompper2020}, such as stochastic motors \cite{RMP1997,pumm2022dna} and microswimmers \cite{Golestanian2009}, is of fundamental importance to the study of biological and synthetic active matter \cite{nardini2017entropy,pietzonka2017entropy,shankar2018hidden,battle2016broken}. These systems typically produce net motion or mechanical work as a consequence of the dissipation of some form of locally available energy (e.g.~ATP hydrolysis) \cite{mugnai2020theoretical,borsley2022chemical,Golestanian2019phoretic}. A common assumption in the literature is that the dissipation can be quantified by representing the autonomous self-propulsion via an effective external ``active force'' \cite{Speck2016,shankar2018hidden,Szamel2019,Tociu.Vaikuntanathan2019,Dabelow.Eichhorn2019}.

The thermodynamic uncertainty relation (TUR) and its various generalizations quantify the trade-off between the precision of a non-equilibrium current and its associated dissipation, and thus provide a powerful tool to infer the underlying driving forces of a system from experimental measurements of its trajectories \cite{barato,gingrich2016dissipation,godec,seifert2019stochastic}. However, many of the models studied until now to test the behavior of TURs have been based on the assumption of tight coupling between chemical and spatial degrees of freedom (where e.g.~a chemical reaction always corresponds to a mechanical step and vice versa) so that spatial, chemical, and entropy production currents effectively become one and the same \cite{barato,pietzonka2016universal,barato2016cost,lee2018thermodynamic,marsland2019thermodynamic,koyuk2022thermodynamic,leighton2022dynamic}.

While convenient, these two (related) assumptions are not generally valid. Realistic autonomous swimmers and molecular motors involve at least two distinct currents (e.g.~spatial and chemical) and consequently at least two distinct kinds of driving forces. In fact, the relevant coupling in these systems is \emph{off-diagonal} (in the language of linear response), as chemical forces drive motion. This is particularly evident in the case of microswimmers, where the force-free constraint on their self-propulsion mechanisms introduces additional complexities \cite{purcell1977life,najafi2004simple,golestanianpre,prl2008,lauga2009hydrodynamics,golestanian2005propulsion,julicher2009generic,gaspard2018fluctuating}, and has important consequences on the bounds on entropy production \cite{nasouri2021minimum,daddi2023minimum}. A multidimensional version of the TUR (MTUR) can in principle be used in multi-current systems to obtain much-improved bounds on the entropy production, and thus better inference of the underlying driving forces \cite{dechant}. However, how to exploit this bound in practice is unclear, as typically only the spatial current is measurable, while the chemical current is not. To shed light on the inner workings of autonomous swimmers and motors, we must therefore understand how spatial and chemical forces and currents couple to each other arbitrarily far from equilibrium, beyond linear response.

\begin{figure}[b]
	\centering
	\includegraphics[width=.8\linewidth]{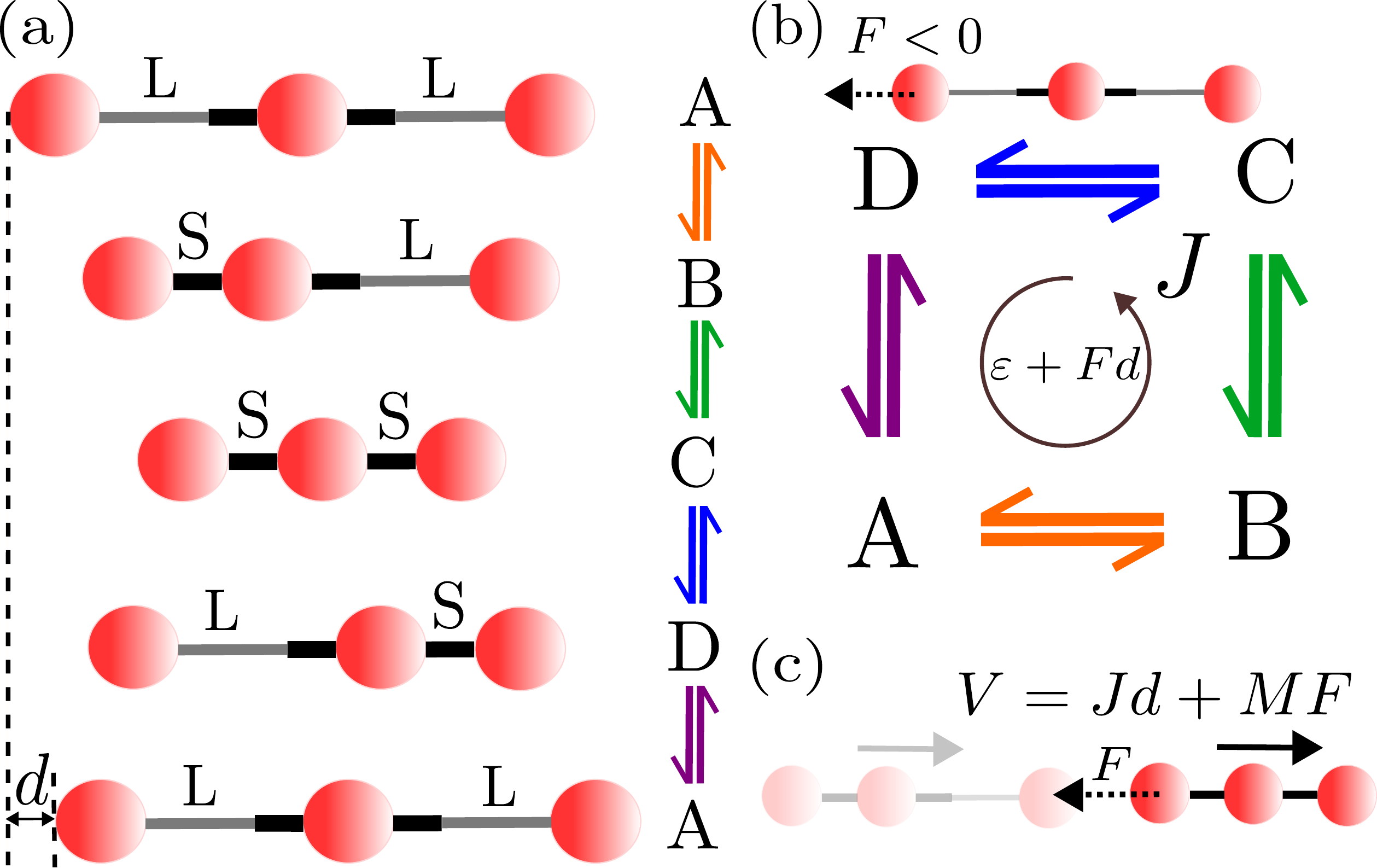}
	\caption{(a) Full four-state cycle of the stochastic three-sphere swimmer, after which it advances a distance $d$. L (long) and S (short) indicate the state of the arms.
	(b) The current $J$ through the cycle is driven by the total affinity $\epsilon+Fd$, where $\epsilon$ is the chemical affinity and $F$ the external force applied on the swimmer. (c) The total velocity of the swimmer $V$ includes an active swimming contribution $Jd$ and a passive drag $M F$, with $M$ the hydrodynamic mobility.}
	\label{fig:intro}
\end{figure}

In this Letter, we study a stochastic three-sphere swimmer \cite{prl2008} as a minimal model that includes both chemical and spatial (hydrodynamic) degrees of freedom; see Fig.~\ref{fig:intro}. The chemical cycle is represented by a four-state  process where each state corresponds to a different conformation of the swimmer. The key hydrodynamic degree of freedom corresponds to the spatial position of the swimmer, to which an external force may also be applied. The total chemical energy $\epsilon$ associated with a cycle and the external force $F$ are the two \emph{affinities} that drive the system out of equilibrium and cause the overall swimming. From a hydrodynamic derivation, we show that the entropy production rate (EPR) can be written as
\begin{align}\label{eq:epr}
    T \dot \sigma = J(\epsilon, F) \epsilon + V(\epsilon, F) F,
\end{align}
where $J(\epsilon,F)$ is the chemical current (rate) of the internal cycle and $V(\epsilon, F)$ is the velocity of the swimmer. Superficially, Eq.~(\ref{eq:epr}) appears to have the standard form of an EPR, with the chemical current $J$ driven by the chemical affinity $\epsilon$ and the spatial current $V$ driven by the spatial affinity $F$. However, the hydrodynamics of the swimming mechanism leads to a coupling between the chemical and spatial degrees of freedom, such that the currents $J$ and $V$ do not respectively vanish when $\epsilon$ and $F$ vanish. In fact, we find that the velocity of the swimmer is given by
\begin{align}\label{eq:velocity}
    V(\epsilon,F) = J(\epsilon,F) d + M F,
\end{align}
where the first term represents the active swimming (with $d$ being the distance advanced in the lab frame after a full conformational cycle), while the second term is the passive drag of the swimmer by the external force. Here, $M$ is the hydrodynamic mobility of the swimmer, which is related to its positional thermal diffusion coefficient through the fluctuation-dissipation relation $D_\mathrm{th} = M k_B T$, where $k_B$ is the Boltzmann constant and $T$ is the temperature. Introducing (\ref{eq:velocity}) into (\ref{eq:epr}), we can rewrite the EPR as
\begin{align}\label{eq:epr2}
    T \dot \sigma = J(\epsilon, F) (\epsilon+Fd) + MF^2.
\end{align}
While perhaps less intuitive, (\ref{eq:epr2}) can be viewed as the \emph{canonical} form of the EPR. Indeed, we show below that the external force influences the dynamics of the chemical cycle through the swimmer mechanics, such that the overall affinity driving the chemical current $J$ is $\epsilon+Fd$.  The passive drag velocity $M F$ in turn represents a hidden current that is exclusively driven by the external force.
Therefore, although Eq.~\eqref{eq:epr} is more practical as it involves currents that are observable (at least in principle), it is only when written in the form of Eq.~(\ref{eq:epr2}) that the non-negativity of the entropy production rate as required by thermodynamics becomes manifest. Below, we derive these results, and show how they strongly influence the precision-dissipation trade-off for the swimmer, and how they can be used to infer the chemical driving force of the swimmer from measurements of its position only.

\textit{Model.---}The model is summarized in Fig.~\ref{fig:intro}(a). We assume quick expansions or contractions of the arms such that their possible states are contracted ($u_\rho=0$) or expanded ($u_\rho=\delta$), where $u_\ell$ and $u_r$ represent the deformation of the left and right arms, respectively, and $\delta$ is the extension amplitude. Each conformation corresponds to a state $\alpha=A,B,C,D$ of the chemical cycle. The rate for the transition $\alpha \rightarrow \beta$ is denoted as $k_{\beta \alpha}$. The states can also be named based on the arms being long ($L$) or short ($S$), e.g.~state $B$ corresponds to $SL$. To introduce forward propulsion (towards the right) without an external force, one must break detailed-balance such that the the trajectory of the system follows closed cycles in the conformational space \cite{najafi2004simple,prl2008,golestanianpre}. If the energy released in going from $\alpha$ to $\beta$ is $\Delta \epsilon_{\beta \alpha}$, local detailed-balance requires $k_{\beta \alpha}/k_{\alpha \beta} = e^{\Delta \epsilon_{\beta \alpha}/k_B T}$. The total affinity of a cycle is then given by 
\begin{math} \displaystyle
\epsilon = \sum \Delta\epsilon_{\beta \alpha} = k_B T \ln{\frac{k_{BA}k_{CB}k_{DC}k_{AD}}{k_{AB}k_{BC}k_{CD}k_{DA}}}.
\end{math}
The steady-state probability current $J$ is given by
$J = k_{BA} P_A - k_{AB} P_B$, where $P_\alpha$'s are the steady-state probabilities \cite{suppmat}.
This current can be viewed as the rate or the inverse period for completion of a chemical cycle. For $\epsilon=0$, equilibrium is restored and the current vanishes.

\textit{Effect of an external force.---}We now consider that the swimmer is pulled or pushed by a constant external force $F$, with the convention that negative force points towards the left (against the direction of swimming). In principle, the force could be distributed among the three spheres such that $F_1+F_2+F_3 = F$ where force $F_i$ is applied on the $i$th sphere. Importantly, the force not only directly drags the swimmer, but also affects its conformational dynamics by modifying the transition rates $k_{\beta \alpha}$. From a hydrodynamic derivation of entropy production (Appendix A), we calculate the total dissipation per transition which includes the work done by the external forces due to the displacements of each sphere. 
\begin{table}
  \begin{center}
    \begin{tabular}{cccc} 
     \hline \hline 
     Process & $\Delta x_{1,\beta \alpha}/\delta$ & $\Delta x_{2,\beta \alpha} /\delta$ & $\Delta x_{3,\beta \alpha}/\delta$\\
      \hline 
      $\quad A=LL \longrightarrow B=SL \quad $ & $1-\alpha_L $  & $\quad -\alpha_L \quad$ & $-\alpha_L$\\ 
      $\quad B=SL \longrightarrow C=SS \quad $  & $\quad \alpha_S \quad$ & $\alpha_S$ & $-(1-\alpha_S)$\\
      $\quad C=SS \longrightarrow D=LS \quad $ & $-(1-\alpha_S)$ & $ \quad \alpha_S \quad$ & $\alpha_S$\\
      $\quad D=LS \longrightarrow A=LL \quad $ & $-\alpha_L$ & $\quad -\alpha_L \quad$ & $1-\alpha_L$\\
      \hline \hline 
    \end{tabular}
  \end{center}
  \caption{Displacement of each sphere in each transition (see Appendix A). For the reverse transitions, $\Delta x_{i,\alpha \beta} = -\Delta x_{i,\beta \alpha}$. The constants $\alpha_{L,S}$ depend on the geometry of the swimmer and satisfy $\frac{1}{3} < \alpha_L < \alpha_S < \frac{1}{2}$. The total displacement $d$ of the swimmer after a full cycle, obtained by summing over any of the columns, is $d=2(\alpha_S-\alpha_L)\delta$. \label{tab:table}}
\end{table}

\begin{figure*}[t]
	\centering
	\includegraphics[width=1\linewidth]{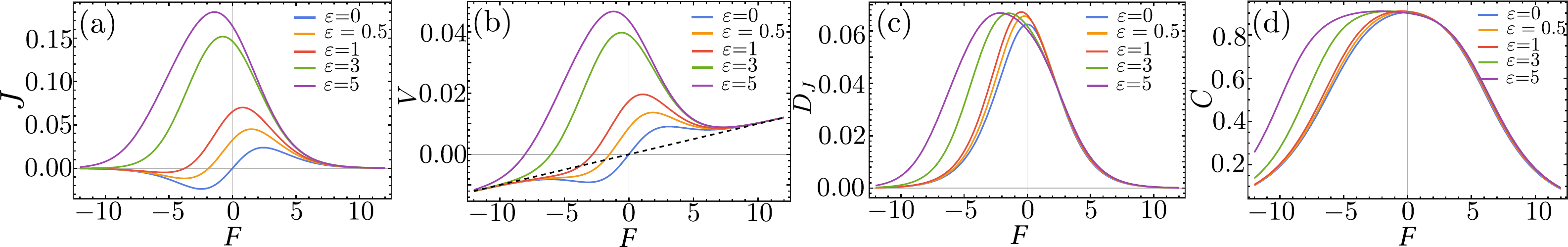}
	\caption{Dependence on external force $F$, for several values of the chemical affinity $\epsilon$, of (a) the chemical current $J$, (b) the spatial current or velocity $V$, (c) the chemical diffusion coefficient $D_J$, and (d) the correlation $C$ between $J$ and $V$. In (b), the dotted black line represents the passive drag velocity $V=MF$.}
	\label{fig:velocity}
\end{figure*}

 The displacements $\Delta x_{i,\beta \alpha}$ of the $i$th sphere during transition $\alpha \rightarrow \beta$ are listed in Table~\ref{tab:table}. Denoting the rates in the absence of the external forces (or the bare rates) as $k_{0\beta \alpha}$, local detailed-balance demands that the rates be modified as
\begin{align}\label{eq:rates}
    k_{\beta \alpha} = k_{0\beta \alpha} \exp\left( \frac{\theta_{\beta \alpha}W_{\beta \alpha}}{k_B T}\right) ,
\end{align}
where $W_{\beta \alpha} = \sum_i F_i \Delta x_{i,\beta \alpha}$ is the work done by the external forces. Since the signs of the displacements are reversed in the reverse transitions, this implies that  $W_{\alpha \beta }=-W_{\beta \alpha }$. 
The factors $\theta_{\beta \alpha}$ are related to the location of the energy barrier between states $\alpha$ and $\beta$, and must satisfy $\theta_{\alpha \beta } = 1 - \theta_{ \beta \alpha}$. Importantly, independently of the choice of $\theta_{\beta \alpha}$ and of where the force is applied, the total affinity of the cycle becomes
\begin{math}\displaystyle
k_B T \ln{\frac{k_{BA}k_{CB}k_{DC}k_{AD}}{k_{AB}k_{BC}k_{CD}k_{DA}}} = \epsilon+Fd.
\end{math}
Using these ingredients and existing results from the literature \cite{derrida,koza} we calculate the chemical current $J$ and its associated diffusion coefficient $D_J$ \cite{suppmat}. Because each conformational cycle results in a displacement $d=2(\alpha_S-\alpha_L)\delta$ (see Table~\ref{tab:table}), the active swimming contributes $J d$ to the velocity $V$ of the swimmer, while the force $F$ additionally contributes a passive drift; see Eq.~\eqref{eq:velocity}. The hydrodynamic mobility $M$ is to the leading order constant during the whole cycle, and more generally, it is an average over all conformations of the swimmer \cite{suppmat}. The spatial diffusion coefficient (associated to $V$) also includes active swimming and passive hydrodynamic contributions, and reads \cite{suppmat}
\begin{equation}
    D_V = D_J d^2 + D_\mathrm{th}. \label{eq:DV}
\end{equation}
Using Eq.~\eqref{eq:velocity}, we can also calculate the correlation between $J$ and $V$ as \cite{suppmat}
\begin{equation}
    C =1/\sqrt{1+D_\mathrm{th}/(D_J d^2)}. \label{eq:C}
\end{equation}

\textit{Swimmer dynamics.---}In all of the following results, we fix the geometric parameters to $\alpha_L=\frac{2.1}{6}$ and $\alpha_S=\frac{2.9}{6}$; the force is applied on the leftmost (trailing) sphere so that $F_1=F$ and $F_2=F_3=0$; and we set $\theta_{\beta \alpha}=1/2$ for all the transitions. The bare transition rates $k_{0 \beta \alpha}$ are all set to the same value $k$, with the exception of $k_{0BA}$ which is set to $k_{0BA}=k e^{\epsilon/k_B T}$. Furthermore, we focus on strong swimmers and set $D_\mathrm{th}/k\delta^2=10^{-3}$. In all plots, quantities are non-dimensionalized using $k^{-1}$ as the timescale, $\delta$ as the length scale, and $k_B T$ and $k_B$ as units of energy and entropy, respectively.

Figure~\ref{fig:velocity} displays the behavior of several quantities of interest as a function of the applied force $F$, for various values of the chemical affinity $\epsilon$. The chemical current, shown in Fig.~\ref{fig:velocity}(a), clearly manifests the mechanochemical coupling in this system, as an applied force can create a chemical current even in the absence of any chemical driving ($\epsilon=0$). When $\epsilon>0$, the current vanishes at the critical force $F_*=-\epsilon/d$ that makes the total affinity of the cycle zero, while it is reversed for $F < F_*$. The behavior of the velocity $V$ is similar to that of $J$ [see Fig.~\ref{fig:velocity}(b)] except that $V$ shows a linear dependence as $V=MF$ at large force, when it is dominated by the passive drag by the external force. The velocity vanishes at the stall force $F_\mathrm{s}$, which can be calculated from the implicit equation $J(\epsilon,F_\mathrm{s})d+MF_\mathrm{s}=0$ [see Eq.~\eqref{eq:velocity}] and satisfies $F_*<F_\mathrm{s}<0$ (for $\epsilon>0$). For sufficiently large $\epsilon$, small positive forces cause the swimmer to decelerate, whereas small negative forces cause the swimmer to accelerate \cite{prl2008}. This phenomenon, known as \emph{negative differential mobility}, has also been observed in other non-equilibrium systems \cite{zia2002getting,benichou2014microscopic}. Lastly, the force dependence of the chemical diffusion coefficient $D_J$ (which coincides with that of the spatial diffusion $D_V$, except for a prefactor and a constant baseline) and the correlation $C$ are displayed in Fig.~\ref{fig:velocity}(c,d). We find that both generally peak at small negative values of the applied force, independently of the magnitude of $\epsilon$.

\begin{figure*}
\centering
\includegraphics[width=1\linewidth]{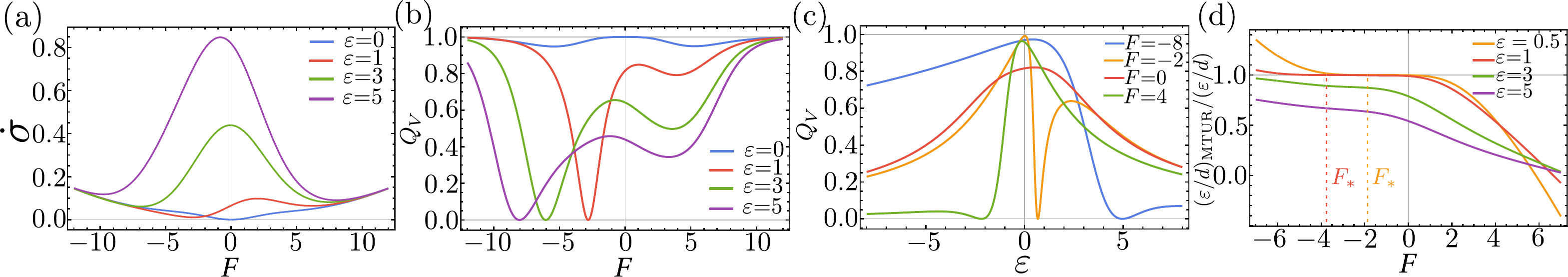}
\caption{(a) Entropy production rate $\dot{\sigma}$ as a function of the external force $F$, for several values of the chemical affinity $\epsilon$. (b,c) The quality factor $Q_V$ of the precision-dissipation trade-off for the spatial current $V$, (b) as a function of $F$ for fixed values of $\epsilon$, (c) as a function of $\epsilon$ for fixed values of $F$. (d) Ratio of the MTUR-inferred [Eq.~\eqref{eq:inference}] and true chemical force $\epsilon/d$. $F_*$ is the critical force at which the chemical current vanishes. For $F>F_*$ the inferred value is a lower bound, for $F<F_*$ an upper bound, for $F=F_*$ it is exact.}
\label{fig:tur}
\end{figure*}

\textit{Entropy production.---}We can calculate the EPR from the hydrodynamic definition of dissipation, $T \dot \sigma =  \sum_i \langle\dot x_i f_i\rangle$, where $\dot x_i$ is the velocity of each sphere and $f_i$ is the corresponding instantaneous force, satisfying the force balance $f_1+f_2+f_3 = F$.
The crucial step in the derivation is to separately consider the internal and external contributions to the forces (see Appendix A). In this framework, the EPR splits into an active swimming contribution related to the conformational transitions, and a purely passive one. After averaging, these two contributions make up the result presented in Eq.~\eqref{eq:epr2}. The coupling between chemical and hydrodynamic driving forces gives rise to a rather complicated dependence of the EPR on the forces, with local maxima and minima as shown in Fig.~\ref{fig:tur}(a). For $\epsilon=F=0$, the system is at equilibrium. For $\epsilon>0$, at low force the entropy production is largely dominated by the chemical part, while at large force we recover the usual hydrodynamic energy dissipation ($\sim F^2)$ of a passive object dragged by a constant force.

\textit{Thermodynamic precision.---}The precision of a non-equilibrium process is bounded by the EPR through the TUR \cite{barato}. 
More specifically, the MTUR provides the bound $\bm{\mathcal{J}}^T \cdot\bm{\mathcal{D}}^{-1} \cdot \bm{\mathcal{J}} \leq  \dot \sigma/k_B$ at steady-state, where $\bm{\mathcal{J}}$ is any vectorial current and $\bm{\mathcal{D}}$ is the diffusion matrix describing the fluctuations of the current \cite{dechant}. Applying this bound to the individual current $V$, we obtain the standard TUR, $ V^2/D_V \leq \dot \sigma/k_B$. We estimate the quality of this bound using the (non-negative) factor $Q_V \equiv (V^2/D_V) /(\dot \sigma/k_B)$, which equals one when the bound is saturated and is smaller otherwise. The values of $Q_V$ for our swimmer are shown in Fig.~\ref{fig:tur}(b) as a function of the force $F$ for several values of the chemical driving $\epsilon$, and in Fig.~\ref{fig:tur}(c) as a function of $\epsilon$ for several values of $F$. The behavior of $Q_V$ is rather complex, reflecting the fact that the current has two driving forces which can compete with each other.

An intriguing observation can be made by considering the behavior of $Q_V$ near equilibrium ($\epsilon=F=0$). While in the limit $F\to 0$ for $\epsilon=0$ in Fig.~\ref{fig:tur}(b) we find $Q_V \to 1$ (the bound is saturated), in the limit $\epsilon \to 0$ for $F=0$ in Fig.~\ref{fig:tur}(c) we find $Q_V \to 0.8163$ (the bound is not saturated). The two limits do \emph{not} coincide as one might have naively expected, implying that the near equilibrium limit is not uniquely defined. This generic behavior can be understood in linear response (see Appendix B), and reflects the fact that $F$ is the direct (diagonal) driving force of the current $V$, whereas $\epsilon$ is its indirect (off-diagonal) driving force \cite{note1}.  This difference between direct and indirect forces with regards to TUR saturation near equilibrium is of practical relevance, as typically (e.g.~for molecular motors and swimmers) the affinity of interest (e.g.~ATP hydrolysis) only indirectly drives the measurable current (e.g.~spatial velocity).

A tighter bound on entropy production can be obtained by applying the MTUR to the two-dimensional current $(J,V)$, which yields
\begin{equation}
\frac{1}{1-C^2}\bigg[ \frac{J^2}{D_J} - \frac{2 C J V}{\sqrt{D_J D_V}}+\frac{V^2}{D_V}\bigg]\leq \frac{\dot \sigma}{k_B }\label{eq:mtur}.  
\end{equation}
This bound is well-behaved and saturated in the near equilibrium limit, as can be proven in the linear response regime (see Appendix B). While it may appear to be less useful in practice, as it involves $J$, $D_J$, and $C$, all of which are not directly observable in an experiment that only has access to the swimmer position, our knowledge of the swimmer mechanics can be exploited to obtain a much improved bound (even an equality) on the chemical energy consumption per unit distance (\emph{chemical force}) of the swimmer, $\epsilon/d$.

\textit{Thermodynamic inference.---}Indeed, using Eqs.~\eqref{eq:velocity}, \eqref{eq:DV}, and \eqref{eq:C}, we can express $J$, $D_J$, and $C$ as functions of $V$, $D_V$, the passive hydrodynamic mobility $M$ (or thermal diffusion coefficient $D_\mathrm{th}=Mk_B T$), and the external force $F$; substitute them into \eqref{eq:mtur} together with the expression \eqref{eq:epr2} for the EPR; and finally rearrange the terms to obtain an inequality on $\epsilon/d$. Defining
\begin{equation}
    \left(\frac{\epsilon}{d}\right)_\mathrm{MTUR} \equiv \frac{k_B T V/D_V-F}{1-D_\mathrm{th}/D_V}, \label{eq:inference}
\end{equation}
we find that, for $F>F_*$ (where $F_*<0$ is the critical force at which the chemical current vanishes), the MTUR provides a lower bound $\epsilon/d \geq (\epsilon/d)_\mathrm{MTUR}$, while for $F<F_*$ the MTUR provides an upper bound $\epsilon/d \leq (\epsilon/d)_\mathrm{MTUR}$. By continuity, for $F=F_*$ [which implies $J=0$ and thus $V=MF_*$ through Eq.~\eqref{eq:velocity}], we find the equality $\epsilon/d = (\epsilon/d)_\mathrm{MTUR} = -F_*$. This result is consistent with the total affinity $\epsilon + Fd$ of the chemical current vanishing at $F=F_*$. The quality of this bound as measured by the ratio $(\epsilon/d)_\mathrm{MTUR}/(\epsilon/d)$ is shown in Fig.~\ref{fig:tur}(d), as a function of $F$, for several values of $\epsilon$.

These results reveal several strategies for the inference of the chemical force. In passive measurements with $F=0$, we can estimate $\epsilon/d \geq k_B T (V/D_V)(1-D_\mathrm{th}/D_V)^{-1}$. This improves the bound obtained from the standard TUR
by a factor $(1-D_\mathrm{th}/D_V)^{-1}$. In active measurements with $F \neq 0$, one may measure the stall force of the swimmer $F_\mathrm{s}<0$ at which $V=0$, as well as the position fluctuations giving $D_V$, and infer $\epsilon/d \geq - F_\mathrm{s} (1-D_\mathrm{th}/D_V)^{-1}$. Lastly, and optimally, one may measure the critical force $F_*$ at which the swimmer moves precisely at the velocity that one would expect from passive hydrodynamic drag, i.e.~$V=MF_*$. The chemical force is then exactly $\epsilon/d = -F_*$. In all these cases, the passive mobility $M$ (and associated $D_\mathrm{th}=Mk_B T$) may be estimated from purely hydrodynamic calculations, or measured from the asymptotic behavior of the swimmer velocity at high force $F\to \pm \infty$; see the force-velocity ($F$-$V$) curve in Fig.~\ref{fig:velocity}(b). In practice, $F_*$ and thus $\epsilon/d$ could be inferred by measuring the $F$-$V$ curve, estimating its high-force asymptote, and obtaining the intersection point of the asymptote and the $F$-$V$ curve, which occurs at $F_*$.

\textit{Discussion---}Using a stochastic three-sphere swimmer as an analytically tractable and thermodynamically-consistent model for a chemically-powered autonomous microswimmer, we have explicitly calculated its swimming dynamics and its entropy production in the presence of an external force. This allowed us to study the coupling between spatial and chemical forces and currents beyond the linear response regime. We have found a number of interesting properties in the force-dependence of the swimmer dynamics as well as its thermodynamic properties, such as the radically different dependence of the thermodynamic precision-dissipation trade-off of the swimmer velocity on the external force (direct driving) and internal chemical affinity (indirect driving). Moreover, we have shown how the chemical affinity of the swimmer can be precisely inferred by measurements of the spatial dynamics only.

All of these properties are a consequence of the coupling between spatial and chemical forces and currents in this system, as exemplified by Eqs.~\eqref{eq:epr}, \eqref{eq:velocity}, \eqref{eq:epr2}, \eqref{eq:DV}, and \eqref{eq:C}. Importantly, we expect the form of these equations to remain unchanged for other autonomous swimmers (e.g.~self-phoretic swimmers \cite{golestanian2005propulsion,julicher2009generic,gaspard2018fluctuating,de2022onsager}) and molecular motors. Only the precise functional form of the chemical current $J(\epsilon,F)$ (and associated diffusion $D_J$) and the distance $d$ advanced per cycle will depend on the microscopic details. Therefore, the general lessons learned here and the thermodynamic inference strategy proposed [Eq.~\eqref{eq:inference}] should be applicable to a wide range of autonomous motors and swimmers.

 \textit{Appendix A: Hydrodynamic calculation of velocity and entropy production---}
The force balance equation for the swimmer reads $\sum_i f_i =F$. The force on sphere $i$ denoted as $f_i$ can be expressed as $f_i=  \tilde f_i+F_i$, in terms of the internal force $\tilde f_i$ and the external force $F_i$. We have $\sum_i \tilde f_i = 0$ and $\sum_i F_i =F$. The internal forces can be written as $\tilde f_1 = -\tilde f_\ell$, $\tilde f_2 = \tilde f_\ell - \tilde f_r$, and $\tilde f_3 = \tilde f_r$, where $\tilde f_\ell$ and $\tilde f_r$ are respectively the internal stresses on the left and right arms, defined to be positive when they act to expand the arm. We denote the instantaneous speeds of the arms as $\dot u_\ell = \dot x_2 -\dot x_1$, $\dot u_r =\dot x_3 - \dot x_2$. Forces and velocities are linearly related through the hydrodynamic friction tensor $Z_{ij}$ such that $f_i = \sum_j Z_{ij} \dot x_j$. By summing over the forces we deduce that $\sum_i A_i \dot x_i = B F$, with  $A_i(u_\ell,u_r) \equiv \left(\sum_j Z_{ij}\right)/\left(\sum_{i,j} Z_{ij}\right)$ and $B(u_\ell,u_r) \equiv 1/\left(\sum_{i,j} Z_{ij}\right)$. From this expression, we derive equations describing the motion of each of the spheres in response to the external force as well as the changes in the lengths of the arms (active swimming) as follows
\begin{align}
    \dot x_1 &= BF -(1-A_1)\dot u_\ell - A_3 \dot u_r ,\label{eq:x1}\\
    \dot x_2 &= BF + A_1 \dot u_\ell - A_3 \dot u_r ,\\
    \dot x_3 &= BF + A_1 \dot u_\ell + (1-A_3)\dot u_r .\label{eq:x3}
\end{align}

The average velocity $V$ of the swimmer is given by $\langle \dot{x}_i \rangle$, where the average is taken at steady state and any sphere $i$ may be considered without loss of generality. Choosing $i=2$, we can write
\begin{align}
\label{eq:veloc1}
   V = \langle \dot x_2 \rangle = M F +\langle A_1 \dot u_\ell - A_3 \dot u_r \rangle
\end{align}
where we have defined $M \equiv \langle B \rangle$, which is independent of $\epsilon$ or $F$ to leading order in the hydrodynamic interactions \cite{suppmat}. The first term represents the passive drag, whereas the second term represents the active swimming. The latter results in finite contributions for each conformational change, so that
\begin{align}\label{eq:exampleofswimming}
\langle A_1 \dot u_\ell - A_3 \dot u_r \rangle &= \frac{1}{\mathcal{T}}\int^{\mathcal{T}}_0 (A_1 \dot u_\ell - A_3 \dot u_r )dt, \nonumber \\ 
&= J \sum_{\{\beta\alpha\}} \left[ \int_\alpha^\beta A_1 du_\ell - \int_\alpha^\beta A_3 du_r\right], \nonumber\\
&= J \sum_{\{\beta\alpha\}} \Delta x_{2,\beta\alpha} =Jd,
\end{align}
where $\mathcal{T}=J^{-1}$ is the period of a cycle in steady state, and the sums run over the forward transitions $\{\beta\alpha\}=\{BA, CB, DC, AD\}$. Equations (\ref{eq:veloc1}) and (\ref{eq:exampleofswimming}) together result in Eq.~(\ref{eq:velocity}).

The displacement in each transition $\Delta x_{2,\beta\alpha}$ is calculated by performing the associated integral in the second line of Eq.~(\ref{eq:exampleofswimming}). For instance, in the transition $A \rightarrow B$ the left arm shrinks and the right arm stays fixed at $u_r = \delta$, so that we find
 \begin{align}\label{eq:deltax2}
    \Delta x_{2,BA} = \int_A^B A_1 du_\ell = \int_\delta^0 A_1(u_\ell,\delta)  d u_\ell = - \alpha_L \delta,
\end{align}
where we have defined
 \begin{align}\label{eq:alphaL}
    \alpha_L \equiv \frac{1}{\delta} \int_0^{\delta}A_1(u_\ell,\delta)  d u_\ell.
\end{align}
Conversely, in the transition $B \to C$, the right arm shrinks while the left arm stays fixed at $u_\ell=0$. We thus find
 \begin{align}\label{eq:deltax2b}
    \Delta x_{2,CB} = - \int_B^C A_3 du_r = - \int_\delta^0 A_3(0,u_r)  d u_r = \alpha_S \delta,
\end{align}
with the definition
 \begin{align}\label{eq:alphaS}
    \alpha_S \equiv \frac{1}{\delta} \int_0^{\delta}A_3(0,u_r)  d u_r.
\end{align}
Repeating this procedure for all transitions and calculating the corresponding displacement of the first and third sphere, we obtain the results in Table~\ref{tab:table}, which are valid for a swimmer with symmetric geometry as in Fig.~\ref{fig:intro}. The calculations for a more general asymmetric swimmer, their explicit integration using the Oseen approximation, and the estimation of the bound $1/3<\alpha_L < \alpha_S < 1/2$ are performed in Ref.~\cite{suppmat}. These recover the known results relating the velocity of the swimmer to the area swept by cycles in conformational space \cite{golestanianpre,prl2008}.

To calculate the EPR at steady state, we start from the hydrodynamic dissipation $T \dot \sigma =  \sum_i  \langle\dot x_i f_i\rangle$. 
Using (\ref{eq:x1})--(\ref{eq:x3}), the EPR becomes
\begin{align}\label{eq:explicitepri}
    T \dot \sigma = M F^2 + \langle \dot u_\ell [\tilde f_\ell - (1-A_1) F_1 + A_1(F_2+F_3)] + \nonumber\\ \dot u_r [\tilde f_r - A_3(F_1+F_2) + (1-A_3) F_3] \rangle.
\end{align}
Here, analogously to the calculation of the velocity, the first term represents the dissipation due to passive drag, whereas the second term represents dissipation due to the active swimming, which gives a finite contribution for each conformational transition.

Indeed, following the same procedure used to derive Eq. \eqref{eq:exampleofswimming}, we may write
\begin{align}\label{eq:eprapp}
    T \dot \sigma = M F^2 + J  \sum_{\{\beta\alpha\}} T \Delta \sigma_{\beta\alpha}, 
\end{align}
where $ T \Delta \sigma_{\beta\alpha}$ gives the dissipation occurring during the transition $\alpha \to \beta$. As an example, we consider again the transition $A \rightarrow B$. In this transition, $u_\ell$ shrinks whereas $u_r$ remains constant and fixed to $\delta$. Thus, the discrete dissipation during this transition is
\begin{align}
  T\Delta \sigma_{BA} &=   \int^0_\delta [\tilde f_\ell -(1-A_1)F_1 + A_1(F_2+F_3) ]du_\ell \nonumber\\ &=\Delta \epsilon_{BA}+F_1 \Delta x_{1,BA} + F_2 \Delta x_{2,BA} + F_3 \Delta x_{3,BA} .
\end{align}
Here, the first term results from the definition
\begin{equation}
\Delta \epsilon_{BA} \equiv  \int^0_\delta \tilde f_\ell du_\ell,
\end{equation}
and gives the dissipation due to the internal active mechanism of the swimmer, while the remaining terms result from the displacements of the spheres previously calculated and give the dissipation by the external force during a conformational change.

An analogous result $T\Delta \sigma_{\beta\alpha} =  \Delta \epsilon_{\beta\alpha}+\sum_i F_i \Delta x_{i,\beta\alpha}$ is obtained for all other transitions \cite{suppmat}. The transition-induced dissipation obtained in this way must be used to enforce local detailed balance in the stochastic transitions, leading to Eq.~(\ref{eq:rates}). Finally, using this result in Eq.~(\ref{eq:eprapp}), and noting that $\sum_{\{\beta\alpha\}} \Delta \epsilon_{\beta\alpha} = \epsilon$ and $\sum_{\{\beta\alpha\},i} F_i \Delta x_{i,\beta\alpha} = F d$, leads to the expression for the EPR in Eq.~(\ref{eq:epr2}).

\textit{Appendix B: Linear response regime---}In the linear-response regime ($\epsilon,Fd\ll k_B T)$, the currents can be written as $J=L_{\epsilon \epsilon} \epsilon + L_{\epsilon F} F$ and $V=L_{F \epsilon} \epsilon + L_{F F} F$, with $L_{\epsilon \epsilon} = \tilde{\kappa} $, $L_{FF}=\tilde{\kappa} d^2 + M $, and $L_{\epsilon F}=L_{F \epsilon }= \tilde{\kappa} d $, where $\tilde{\kappa}=\kappa_0/(k_BT)$ and $\kappa_0$ is an inverse timescale that depends only on the force-free rates $k_{0\beta\alpha}$ \cite{suppmat}. Thus, affinities couple to the currents through $\mathcal{J}_a = \sum_b L_{a b} \mathcal{A}_b$, where $\mathcal{J}_a=(J,V)$ is a vector of currents, $\mathcal{A}_a=(\epsilon, F)$ is a vector of affinities, and $L_{a b}$ is a symmetric Onsager matrix. The EPR can then be expressed in the usual bilinear form $T \dot \sigma = \sum_{a b} L_{a b} \mathcal{A}_a \mathcal{A}_b $ \cite{degroot}.

It is straightforward to show that the MTUR ($\bm{\mathcal{J}}^T \cdot \bm{\mathcal{D}}^{-1} \cdot \bm{\mathcal{J}} \leq  \dot \sigma/k_B$) is saturated in linear regime. Indeed, the diffusion matrix $\bm{\mathcal{D}}$ relates to the Onsager mobility matrix $\bm{L}$ through the fluctuation-dissipation theorem $\bm{\mathcal{D}}=k_B T \bm{L}$, and using this together with $\bm{\mathcal{J}}=\bm{L} \cdot \bm{\mathcal{A}}$ directly results in $\bm{\mathcal{J}}^T \cdot \bm{\mathcal{D}}^{-1} \cdot \bm{\mathcal{J}} =  \dot \sigma/k_B$. The behavior of the standard single-current TUR, $Q_a \equiv (\mathcal{J}_a^2/D_a)/(\dot \sigma/k_B)\leq 1$ for  current $\mathcal{J}_a$, is more surprising. In general, for a current $\mathcal{J}_a$ driven by $N$ affinities $\mathcal{A}_b$ with $b=1,...,N$, the quality factor $Q_a$ can be written as
\begin{equation}
    Q_a = \frac{\sum_{b,c}L_{ab}L_{ac} \mathcal{A}_b  \mathcal{A}_c}{L_{aa}\sum_{b,c}L_{bc}\mathcal{A}_b\mathcal{A}_c}
\end{equation}
where we have used the fluctuation-dissipation relation $D_a = \mathcal{D}_{aa}= k_B T L_{aa}$. There are two distinct cases with regards to how the system behaves when only one of the affinities is nonzero: (i) If  $\mathcal{A}_b=0$ for all $b\neq a$ and $\mathcal{A}_a \neq 0$ (weak direct driving), we find $Q_a = 1$, i.e.~the bound saturates; (ii) If $\mathcal{A}_b=0$ for all $b\neq c$ and $\mathcal{A}_c \neq 0$ for some $c\neq a$ (weak indirect driving), we find $Q_a = L_{ac}^2/(L_{aa} L_{cc})$ \cite{note1}. The latter value is guaranteed to be smaller than or equal to one due to the positive semi-definiteness of the Onsager matrix. Typically, it is smaller than one, implying that the TUR bound is not saturated. For our swimmer, we thus have $Q_V \to L_{F\epsilon}^2/(L_{FF}L_{\epsilon\epsilon})$ when $F=0$ and $\epsilon \to 0$, which for the parameters used in Fig.~\ref{fig:tur} gives $Q_V \to 0.8163$.

\acknowledgements
We acknowledge support from the Max Planck School Matter to Life and the MaxSynBio Consortium which are jointly funded by the Federal Ministry of Education and Research (BMBF) of Germany and the Max Planck Society.

\bibliography{arxiv_3ss}

\begin{thebibliography}{44}
\bibitem [{not()}]{note1}%
  \BibitemOpen
  \href@noop {} {}\bibinfo {note} {While the full expression of the TUR in
  linear response was derived in the seminal Ref.~\cite{barato}, this
  particular consequence appears to have been overlooked.}\BibitemShut {Stop}%
\end{thebibliography}

\begin{widetext}
\renewcommand{\thefigure}{S\arabic{figure}}
\setcounter{figure}{0}

\newpage

\setcounter{equation}{0}
\setcounter{figure}{0}
\setcounter{table}{0}
\renewcommand{\theequation}{S\arabic{equation}}
\renewcommand{\thefigure}{S\arabic{figure}}
\renewcommand{\thetable}{S\Roman{table}}

\begin{center}
\begin{title}\centering
\LARGE{\textit{{Supplemental} Material}}
\end{title}
\end{center}

\section{Currents, diffusion coefficients, and correlation}

To derive the expressions for the currents, diffusion coefficients, and the correlation, we define the observables $n$, which counts the number of cycles completed by the internal process, and $y$, the displacement due to the passive drag by the force. The total displacement of the swimmer is given by $x = nd + y$, where $d=2(\alpha_S-\alpha_L) \delta$ as derived in the main text. At steady-state, we have
\begin{align}
    &\lim_{t\rightarrow \infty} \langle n \rangle/t = J,\\
    &\lim_{t\rightarrow \infty} \langle y \rangle/t = M F,\\
    &\lim_{t\rightarrow\infty} \langle x \rangle/t = J d +M F = V,
\end{align}
where the hydrodynamic mobility $M$ of the swimmer is related to its thermal diffusion coefficient through $M=D_\mathrm{th}/k_B T$. We show (below) through a comprehensive hydrodynamic calculation that the mobility is constant to lowest order, and thus independent of the cyclic process $n$. To calculate $J$, we need to solve the master equation
\begin{align}
    \frac{dp_\alpha}{dt} = \sum_{\beta} (k_{\alpha \beta}p_\beta-k_{\beta \alpha}p_\alpha ),
\end{align}
which governs the dynamics of the probability distribution $p_\beta$. At steady-state we denote $p_\beta \rightarrow P_\beta$, and write the probability current as $J = k_{\beta \alpha} P_\alpha - k_{\alpha \beta} P_\beta$. We find \cite{prl2008}
\begin{align}\label{eq:current}
    J = \frac{k_{AD}k_{DC}k_{CB}k_{BA}-k_{AB}k_{BC}k_{CD}k_{DA}}{\sum_{\mathrm{cycl. perm. \ of } \ A,B,C,D} (k_{AD} k_{DC} k_{CB} + k_{AB}k_{BC}k_{CD}+k_{AB}k_{AD}k_{DC}+k_{AD}k_{AB}k_{BC})}.
\end{align}
The diffusion coefficients are defined as
\begin{align}
\lim_{t\rightarrow \infty}\langle (n - \langle n \rangle)^2 \rangle/(2 t) = D_J,\\
\lim_{t\rightarrow \infty}\langle (y - \langle y \rangle)^2 \rangle/(2 t) = D_\mathrm{th},\\
\lim_{t\rightarrow \infty}\langle (x - \langle x \rangle)^2 \rangle/(2 t) = D_V.
\end{align}
The explicit expression for calculating the diffusion coefficient $D_J$ can be found in \cite{koza}.
Using the above equations and the definition of $x$, the spatial diffusion coefficient can be found as
\begin{align}
\lim_{t\rightarrow \infty}\langle (x - \langle x \rangle)^2 \rangle/(2 t) =  D_J d^2 + D_\mathrm{th} = D_V,
\end{align}
since $\langle ny \rangle - \langle n \rangle \langle y \rangle=0$.
The correlation $C$ between the processes $x$ and $n$ and is defined as
\begin{align}
C = \lim_{t\rightarrow \infty} \frac{\langle (x -\langle x \rangle ) (n -\langle n \rangle ) \rangle }{\sqrt{\langle ( x - \langle x \rangle )^2 \rangle \langle ( n - \langle n \rangle )^2 \rangle }},
\end{align}
and can be straightforwardly calculated as
\begin{align}\label{eq:correlation}
C = \frac{1}{\sqrt{1 + \frac{D_\mathrm{th}}{D_J d^2}}}.
\end{align}

\section{Linear response theory}

Expanding the currents $J$ and $V$ around the equilibrium state ($\epsilon=F=0$), we find
\begin{align}
    J(\epsilon,F) &= \frac{\partial J}{\partial \epsilon}\bigg\vert_{\epsilon=F=0}\epsilon + \frac{\partial J}{\partial F}\bigg\vert_{\epsilon=F=0}F, \\
 V(\epsilon,F) &= d\,\frac{\partial J}{\partial \epsilon}\bigg\vert_{\epsilon=F=0}\epsilon + \bigg(M + d\,\frac{\partial J}{\partial F}\bigg \vert_{\epsilon=F=0}\bigg)F.
\end{align}
To verify the Onsager reciprocity condition, we need to show that  
$\frac{\partial J}{\partial F} = d\frac{\partial J}{\partial \epsilon}$ is satisfied.

To this end, we note that 
\begin{align}
J =\kappa \left(1 - \Pi  \frac{k_{\beta \alpha}^\leftarrow}{k_{\alpha \beta}^\rightarrow}\right)= \kappa \Big(1-\exp(-\mathcal{A}/k_B T)\Big) , 
\end{align}
where
\begin{align}
    \kappa = \frac{k_{AD}k_{DC}k_{CB}k_{BA}}{\sum_{\mathrm{cycl. perm. \ of } \ A,B,C,D} (k_{AD} k_{DC} k_{CB} + k_{AB}k_{BC}k_{CD}+k_{AB}k_{AD}k_{DC}+k_{AD}k_{AB}k_{BC})}.
\end{align}
and the cycle affinity $\mathcal{A}$ is defined as
\begin{align}
    \frac{\mathcal{A}}{k_B T} = \ln\bigg(\frac{k_{AD}k_{DC}k_{CB}k_{BA}}{k_{AB}k_{BC}k_{CD}k_{DA}} \bigg).
\end{align}
Thus, close to equilibrium we find
\begin{align}
    J \simeq \kappa_0\frac{\mathcal{A}}{k_B T} = \kappa_0\frac{(\epsilon+Fd)}{k_B T},
\end{align}
where $\kappa_0= \kappa(\epsilon=F=0)$ and $\mathcal{A} = \epsilon + Fd $ is the affinity. 
Therefore, we deduce that Onsager reciprocity is satisfied, and the Onsager coefficients are as given in the main text. Moreover, one can easily check that in the linear regime the entropy production is given as $T \dot \sigma = J \epsilon + V F$.

\section{Hydrodynamic calculations}

The equations for the velocities of each sphere are given in terms of the
hydrodynamic mobility tensor $M_{ij}$, such that $\dot x_i = \sum_j M_{ij} f_j$.
In what follows we assume arbitrary sizes of the three spheres and also different length of the arms where $u_\ell$ and $u_r$ can take maximum expansions of $\delta_\ell$ and $\delta_r$, respectively.

As describe in Appendix A, it is useful to consider the inverse problem $f_i=\sum_j Z_{ij}\dot x_j$, with the friction matrix $Z = M^{-1}$,
\begin{align}
Z=\frac{1}{\det M}
\begin{bmatrix}
    M_{22}M_{33}-M_{23}^2 & M_{13}M_{23} -M_{33}M_{12}& M_{12}M_{23}-M_{13}M_{22} \\
   M_{13}M_{23} -M_{33}M_{12} & M_{11}M_{33}-M_{13}^2 & M_{12}M_{13}-M_{11}M_{23}\\ M_{12}M_{23}-M_{13}M_{22} & M_{12}M_{13}-M_{11}M_{23} & M_{11}M_{22}-M_{12}^2
\end{bmatrix},
\end{align}
where 
\begin{equation}
\det M= M_{11}M_{22}M_{33}-M_{11}M_{23}^2-M_{22}M_{13}^2-M_{33}M_{12}^2+2M_{12}M_{13}M_{23},  
\end{equation}
is the determinant of $M_{ij}$. We thus obtain that
\begin{align}
    \sum_{ij} Z_{ij} \dot x_j = \sum_i f_i = \sum_i F_i = F 
\end{align}
since $\sum f_i =\sum F_i = F$ and $\sum \tilde f_i =0$ as explained in the main letter. Thus, by dividing both sides with $\sum_{ij} Z_{ij}$, and defining 
\begin{equation}
B \equiv \frac{1}{\sum_{ij} Z_{ij}},    
\end{equation}
and 
\begin{equation}
A_i \equiv \frac{\sum_j Z_{ij}}{\sum_{ij}Z_{ij}},     
\end{equation}
we obtain $\sum_i A_i \dot x_i =  BF$. Using $\dot u_\ell = \dot x_2 -\dot x_1$ and $\dot u_r =\dot x_3 - \dot x_2$ in this expression, we find
\begin{align}
   \dot x_1 &= BF -(1-A_1)\dot u_\ell - A_3 \dot u_r, \label{eq:x1_sup}\\
    \dot x_2&= BF + A_1 \dot u_\ell - A_3 \dot u_r, \\
    \dot x_3 &= BF + A_1 \dot u_\ell + (1-A_3)\dot u_r, \label{eq:x3_sup}
\end{align}
where the explicit expressions are given as follows
\begin{align}
    A_1(u_\ell,u_r) = \frac{1}{N}(M_{22}M_{33}-M_{33}M_{12}-M_{22}M_{13}+M_{13}M_{23}+M_{12}M_{23}-M_{23}^2),\\
    A_2(u_\ell,u_r) = \frac{1}{N}(M_{11}M_{33}-M_{33}M_{12}-M_{11}M_{23}+M_{12}M_{13}+M_{13}M_{23}-M_{13}^2),\\
    A_3(u_\ell,u_r) = \frac{1}{N}(M_{11}M_{22}-M_{11}M_{23}-M_{22}M_{13}+M_{12}M_{23}+M_{12}M_{13}-M_{12}^2).
\end{align}
Here $N$ is determined by $A_1+A_2+A_3=1$. The expression for $B$ is
\begin{align}
    B(u_\ell,u_r)=\frac{1}{N} \det M. 
\end{align}

\subsection{Swimming velocity}

The average swimming velocity $V$ is given by $V=\langle \dot x_i \rangle$ where without loss of generality we choose $\langle \dot x_2 \rangle$ with
\begin{align}
    \langle \dot x_2 \rangle = \langle B \rangle F +\langle A_1 \dot u_\ell - A_3 \dot u_r \rangle = M F +\langle A_1 \dot u_\ell - A_3 \dot u_r \rangle
\end{align}
where in the steady-state 
\begin{equation}
M \equiv \langle B \rangle = B(\delta_\ell,\delta_r) P_A + B(0,\delta_r) P_B+B(0,0)P_C +B(\delta_\ell,0)P_D,   \label{eq:mobility-def}
\end{equation}
with the brackets indicating the state of $B$. For the second term we have,
\begin{align}\label{eq:exampleofswimming_sup}
    \langle A_1 \dot u_\ell - A_3 \dot u_r \rangle &= \frac{1}{\mathcal{T}}\int^{\mathcal{T}}_0 (A_1 \dot u_\ell - A_3 \dot u_r )dt = J \sum_{\{\beta\alpha\}} \left[ \int_\alpha^\beta A_1 du_\ell - \int_\alpha^\beta A_3 du_r\right], \nonumber\\
    &= J(\Delta x_{2,BA} + \Delta x_{2,CB}+\Delta x_{2,DC}+\Delta x_{2,DA})=Jd,
\end{align}
where after integrating over each step we obtain the displacement during each transition, and we have used the fact that in the steady-state $J(\epsilon,F) = \frac{1}{\mathcal{T}}$. 

Performing the individual integrals for each transition we find that for the transition $A=LL \to B=SL$ the following result holds
\begin{align}
    \Delta x_{2,BA} = \int^0_{\delta_\ell}A_1(u_\ell,\delta_\ell) du_\ell = -\alpha_{L,\ell} \delta_\ell~~\text{with}~~\alpha_{L,\ell} \equiv \frac{1}{\delta_\ell}\int^{\delta_\ell}_0 A_1(u_\ell,\delta_r) du_\ell.
\end{align}
Due to the constraint of $\dot u_r =0$, we find that $\Delta x_{3,BA}= - \delta_\ell \alpha_{L,\ell}$, and since $-\delta_\ell = \Delta x_{2,BA} - \Delta x_{1,BA}$, then $\Delta x_{1,BA}=(1-\alpha_{L,\ell})\delta_\ell$. 

For the transition $B=SL \to C=SS$, we find
\begin{align}
    \Delta x_{2,CB} = -\int^0_{\delta_r}A_3(0,u_r) du_r = \alpha_{S,r} \delta_r~~\text{with}~~\alpha_{S,r} \equiv \frac{1}{\delta_r} \int^{\delta_r}_0 A_3(0,u_r) du_r,
\end{align}
where $\Delta x_{1,CB} = \delta_r \alpha_{S,r}$ and $\Delta x_{3,CB} = -(1-\alpha_{S,r})\delta_r$.

For the transition $C=SS \to D=LS$, we find
\begin{align}
    \Delta x_{2,DC} = \int_0^{\delta_\ell}A_1(u_\ell,0) du_\ell = \alpha_{S,\ell} \delta_\ell~~\text{with}~~\alpha_{S,\ell}\equiv \frac{1}{\delta_ \ell} \int^{\delta_\ell}_0 A_1(u_\ell,0) du_\ell,
\end{align}
where $\Delta x_{1,DC} =-(1-\alpha_{S,\ell})\delta_\ell $ and $\Delta x_{3,DC} = \delta_\ell \alpha_{S,\ell} $.

Finally, for the transition $D=LS \to A=LL$, we find
\begin{align}
    \Delta x_{2,AD} = -\int_0^{\delta_r}A_3(\delta_\ell,u_r) du_r = - \alpha_{L,r} \delta_r~~\text{with}~~\alpha_{L,r} \equiv \frac{1}{\delta_r} \int^{\delta_r}_0 A_3(\delta_\ell,u_r) du_r,
\end{align}
where $\Delta x_{1,AD} =-\alpha_{L,r}\delta_r $ and $\Delta x_{3,DC} = (1- \alpha_{L,r}) \delta_r$.

The above results are all summarized in Table \ref{tab:tablesi}. By summing any column of the table, we find that the total displacement due to swimming after a full cycle is
\begin{align}
    d = (\alpha_{S,\ell} - \alpha_{L,\ell})\delta_\ell + (\alpha_{S,r} - \alpha_{L,r})\delta_r.
\end{align}

\begin{table} 
  \begin{center}
    \begin{tabular}{cccc} 
     \hline \hline 
     Process & $\Delta x_{1,\beta \alpha}$ & $\Delta x_{2,\beta \alpha} $ & $\Delta x_{3,\beta \alpha}$\\
      \hline 
      $\quad A=LL \longrightarrow B=SL \quad $ & $(1-\alpha_{L,\ell}) \delta_\ell $  & $\quad -\alpha_{L,\ell} \delta_\ell \quad$ & $-\alpha_{L,\ell} \delta_\ell$\\
      $\quad B=SL \longrightarrow C=SS \quad $  & $\quad \alpha_{S,r} \delta_r \quad$ & $\alpha_{S,r} \delta_r$ & $-(1-\alpha_{S,r})\delta_r$\\
      $\quad C=SS \longrightarrow D=LS \quad $ & $-(1-\alpha_{S,\ell}) \delta_\ell$ & $ \quad \alpha_{S,\ell} \delta_\ell \quad$ & $\alpha_{S,\ell} \delta_\ell$\\
      $\quad D=LS \longrightarrow A=LL \quad $ & $-\alpha_{L,r} \delta_r$ & $\quad -\alpha_{L,r} \delta_r \quad$ & $(1-\alpha_{L,r})\delta_r$\\
      \hline \hline 
    \end{tabular}
  \end{center}
  \caption{The displacement of each sphere in each transition for a general asymmetric swimmer. For the reverse transitions, $\Delta x_{i,\alpha \beta} = -\Delta x_{i,\beta \alpha}$ as in the main text. The first index of $\alpha$ coefficients stands for whether the other (fixed) arm is in the long or short and the second one is for the arm that is changing. \label{tab:tablesi}}
\end{table}

In the special case of a symmetric swimmer like in the main text with all spheres identical and thus $M_{11}=M_{22}=M_{33}=\mu$ as well as $\delta_\ell=\delta_r=\delta$, we find $\alpha_{S,\ell} = \alpha_{S,r}=\alpha_S$ and $\alpha_{L,\ell} = \alpha_{L,r}=\alpha_L$ and the table in the main text is recovered, which implies that $d=2(\alpha_S-\alpha_L)\delta$.
Bounds on the possible values of $\alpha_S$ and $\alpha_L$ can be obtained by considering two extreme cases. If in the $L$ state the spheres are very far from each other, hydrodynamic interactions are negligible and then $A_1=A_2=A_3 =1/3$, giving a lower bound of $\alpha_L=1/3$. If in the $S$ state the spheres are so close to each other as to be overlapping, we can transform the three-sphere problem into a two-sphere problem. We take the  $C \to  D$ transition as an example. The internal active force of the left arm is denoted $\tilde{f}$ (positive if it acts to contract the arm). The positions of the spheres then evolve according to $\dot{x}_1 = (-\mu+M_{12}) \tilde f$ and $\dot{x}_2 = (\mu_S - M_{12}) \tilde f$ with $\mu_S$ defined as the effective mobility of the rigid block. Therefore, we can express $\alpha_S$ as
\begin{align}
    \alpha_{S} = \frac{\mu_{S}-M_{12}}{\mu+\mu_{S}-2 M_{12}}.
\end{align}
For the upper bound one can assume that that the effective block with mobility $\mu_S$ becomes a sphere of mobility $\mu$. Hence, in this limit $\alpha_S =1/2$. With these two bounds, we find that overall $1/3<\alpha_L < \alpha_S < 1/2$ as stated in the main text.

\subsection{Perturbative expansion}

We proceed further to find a perturbative expression for the values of the $\alpha$-coefficients, displacement after a full cycle, and average velocity. We use the Stokes law for the friction coefficient and the expression for the Oseen tensor. For simplicity, we take all spheres to be equal. We assume that $M_{11}=M_{22}=M_{33}\equiv\mu=\frac{1}{6 \pi \eta a}$ where $a$ is the radius of the spheres and $\eta$ is the viscosity of the medium. For the off-diagonal components the expressions are, $M_{12}= \frac{1}{4 \pi \eta s_\ell} $, $M_{23} =\frac{1}{4 \pi \eta s_r}$ and $M_{13} = \frac{1}{4 \pi \eta (s_\ell+s_r)}$, where $s_i$ is the distance between the corresponding spheres. We define, $s_\rho = L_\rho + u_\rho$ with $L_\rho$ being a constant length with $u_\rho \ll L_\rho$. By Taylor expanding, the two integrals in Eq.~\eqref{eq:exampleofswimming_sup} become
\begin{align}
    \int_\mathrm{cycle}A_1 du_\ell = \frac{1}{3}\int_\mathrm{cycle} du_\ell + \frac{a}{3}\int_\mathrm{cycle} \left(K_{0,\ell r} + K_{1,\ell r} u_\ell - K_{2,\ell r} u_r \right)du_\ell+\mathcal{O}\left(\frac{au_\rho^2}{L_\rho^3}\right), \\
    -\int_\mathrm{cycle}A_3 du_r = -\frac{1}{3}\int_\mathrm{cycle} du_r - \frac{a}{3}\int_\mathrm{cycle} \left(K_{0,r \ell} - K_{2,r\ell} u_\ell + K_{1,r \ell} u_r \right)du_r+\mathcal{O}\left(\frac{au_\rho^2}{L_\rho^3}\right),
\end{align}
where
\begin{align}
    K_{0,\rho \nu} &= \frac{L_\rho}{L_\nu(L_\rho+L_\nu)}-\frac{L_\nu}{2 L_\rho(L_\rho+L_\nu)},\\
    K_{1, \rho \nu} &= \frac{1}{(L_\rho+L_\nu)^2}+\frac{L_\nu}{L_\rho(L_\rho+L_\nu)^2}+\frac{L_\nu^2}{2L_\rho^2(L_\rho+L_\nu)^2},\\
    K_{2,\rho \nu} & = \frac{1}{2(L_\rho+L_\nu)^2}+\frac{2L_\rho}{L_\nu(L_\rho+L_\nu)^2}+\frac{L_\rho^2}{L_\nu^2(L_\rho+L_\nu)^2}.
\end{align}
This allows us to calculate the perturbative expressions for $\alpha$ coefficients as follows
\begin{align}
    \alpha_{L,\ell} &= \frac{1}{3} + \frac{a}{3}\left( K_{0,\ell r} + \frac{1}{2} K_{1,\ell r} \delta_\ell - K_{2,\ell r} \delta_r\right),\\
     \alpha_{S,r} &= \frac{1}{3}+\frac{a}{3}\left( K_{0,r \ell} +\frac{1}{2} K_{1,r \ell} \delta_r \right),\\
      \alpha_{S,\ell} &= \frac{1}{3}+\frac{a}{3}\left( K_{0,r \ell} +\frac{1}{2} K_{1,\ell r} \delta_\ell \right),\\
       \alpha_{L,r} &= \frac{1}{3} + \frac{a}{3}\left( K_{0,r \ell} + \frac{1}{2} K_{1,r \ell } \delta_r - K_{2,r \ell } \delta_\ell\right),
\end{align}
which implies
\begin{align}
    d = \frac{a}{3}\left(K_{2,\ell r} + K_{2,r \ell} \right)\delta_\ell \delta r = \frac{a}{3}\left(\frac{1}{L_\ell^2}+\frac{1}{L_r^2}-\frac{1}{(L_\ell+L_r)^2} \right)\delta_\ell \delta r.
\end{align}
This recovers the expected result \cite{prl2008}
\begin{align}
    V(F=0) = Jd = \frac{a}{3}\left(\frac{1}{L_\ell^2}+\frac{1}{L_r^2}-\frac{1}{(L_\ell+L_r)^2} \right) J \delta_\ell \delta r = \frac{a}{3}\left(\frac{1}{L_\ell^2}+\frac{1}{L_r^2}-\frac{1}{(L_\ell+L_r)^2} \right) \left\langle\frac{d\mathcal{A}}{dt}\right\rangle,
\end{align}
for the propulsion of a swimmer in the absence of an external force being proportional to the area $\mathcal{A}$ enclosed by its trajectory in conformation space.

In the special case of $L_\ell=L_r\equiv L_0$, one obtains 
\begin{align}
       \alpha_{L,\ell} &= \frac{1}{3}+ \frac{a}{12L_0^2}\left(L_0+\frac{5}{4}\delta_\ell - \frac{7}{2}\delta_r \right),\\
       \alpha_{S,r} &= \frac{1}{3}+ \frac{a}{12L_0^2}\left(L_0+\frac{5}{4}\delta_r  \right),\\
       \alpha_{S,\ell} &= \frac{1}{3}+ \frac{a}{12L_0^2}\left(L_0+\frac{5}{4}\delta_\ell  \right),\\
       \alpha_{L,r} &=  \frac{1}{3}+ \frac{a}{12L_0^2}\left(L_0+\frac{5}{4}\delta_r-\frac{7}{2}\delta_\ell  \right),
\end{align}
which implies 
\begin{align}
    d = \frac{7a}{12L_0^2}\delta_\ell \delta_r.
\end{align}

If we further assume that $\delta_\ell = \delta_r \equiv \delta$ as in the main text, we find
\begin{align}
       \alpha_{L,\ell} &= \alpha_{L,r} \equiv \alpha_L = \frac{1}{3}+ \frac{a}{12L_0^2}\left(L_0 - \frac{9}{4}\delta \right),\\
       \alpha_{S,\ell} &= \alpha_{S,r} \equiv \alpha_S = \frac{1}{3}+ \frac{a}{12L_0^2}\left(L_0+\frac{5}{4}\delta  \right).
\end{align}
As stated in the main text, we observe that we generically have $\alpha_L<\alpha_S$. The lower bound of $1/3$ corresponds to the limit $L_0 \gg a$. The total displacement after a full cycle is then
\begin{align}
d=\frac{7a}{12L_0^2}\delta^2.
\end{align}
\subsection{Hydrodynamic mobility}
As derived above, the average mobility $M$ describing the passive drag of the particle by the external force is given in Eq. \eqref{eq:mobility-def}.
 
In general, $M$ therefore depends on the chemical affinities and external forces through the probabilities $P_\alpha$, and is not purely geometric. However,
to the first order in the hydrodynamic interactions $\mathcal{O}(M_{ij}/M_{ii})$, the instantaneous mobility $B$ can be written as
\begin{equation}
B = (M_{11}^{-1}+M_{22}^{-1}+M_{33}^{-1})^{-1} \left( 1 + 2 \frac{M_{11}M_{23} + M_{22}M_{13}+M_{33}M_{12}}{M_{11}M_{22}+M_{11}M_{33}+M_{22}M_{33}} + \mathcal{O}\left(\frac{M_{ij}^2}{M_{ii}^2}\right) \right),
\end{equation}
where the dependence on $(u_\ell,u_r)$ is only contained in the cross-mobilities $M_{ij}$ with $i \neq j$, which are of higher order. Therefore, the average mobility $M$ is independent of the chemical affinities and external forces  to the lowest order and reads
\begin{equation}
M = (M_{11}^{-1}+M_{22}^{-1}+M_{33}^{-1})^{-1} + \langle \mathcal{O}(M_{ij}/M_{ii})  \rangle.
\end{equation}
In particular, for equal-sized spheres with mobility $\mu$ we find $M\approx \mu/3$.

\subsection{Entropy production rate}

 As described in Appendix A, starting for the expression of the EPR  $T\dot \sigma = \sum_i \langle \dot x_i f_i \rangle$, splitting the forces into internal and external, and using Eqs.~(\ref{eq:x1_sup})--(\ref{eq:x3_sup}), the EPR becomes
 \begin{align}
    T \dot \sigma = \langle BF^2 + \dot u_\ell [\tilde f_\ell - (1-A_1) F_1 + A_1(F_2+F_3)] +  \dot u_r [\tilde f_r - A_3(F_1+F_2) + (1-A_3) F_3] \rangle.
\end{align}
As explained in the main text the contributions to entropy production from each stochastic transition can be calculated as was done explicitly for the displacements in Eq.~\eqref{eq:exampleofswimming_sup}. By following the procedure per step and averaging over the steady-state limit, one easily finds
\begin{align}
    T \dot \sigma = M F^2 + J(\epsilon,F) \sum_{\{\beta\alpha\}} T\Delta\sigma_{\beta \alpha},
\end{align}
where the contributions to dissipation in each transition are given by the following expressions
\begin{align}
  T\Delta \sigma_{BA} =   \int_{\delta_\ell}^0 [\tilde f_\ell -(1-A_1)F_1 + A_1(F_2+F_3) ]du_\ell  =\Delta \epsilon_{BA}+ \sum_i F_i \Delta x_{i,BA}~\text{with}~\Delta \epsilon_{BA} \equiv \int_{\delta_\ell}^0 \tilde f_\ell du_\ell,
\end{align}
\begin{align}
  T\Delta \sigma_{CB} =   \int_{\delta_r}^0 [\tilde f_r - A_3(F_1+F_2) + (1-A_3) F_3 ]du_r  =\Delta \epsilon_{CB}+ \sum_i F_i \Delta x_{i,CB}~\text{with}~\Delta \epsilon_{CB} \equiv \int_{\delta_r}^0 \tilde f_r du_r,
\end{align}
\begin{align}
  T\Delta \sigma_{DC} =   \int_0^{\delta_\ell} [\tilde f_\ell -(1-A_1)F_1 + A_1(F_2+F_3) ]du_\ell  =\Delta \epsilon_{DC}+ \sum_i F_i \Delta x_{i,DC}~\text{with}~\Delta \epsilon_{DC} \equiv \int_0^{\delta_\ell} \tilde f_\ell du_\ell,
\end{align}
\begin{align}
  T\Delta \sigma_{AD} =   \int_0^{\delta_r} [\tilde f_r - A_3(F_1+F_2) + (1-A_3) F_3 ]du_r  =\Delta \epsilon_{AD}+ \sum_i F_i \Delta x_{i,AD}~\text{with}~\Delta \epsilon_{AD} \equiv \int_0^{\delta_r} \tilde f_r du_r.
\end{align}
By summing the chemical energies $\Delta \epsilon_{\beta \alpha}$ over all four transitions we recover (by definition) the total chemical affinity $\epsilon$. In turn, the sum of $\sum_i F_i \Delta x_{i,\beta\alpha}$ over all four transitions gives $Fd$, as the sum of displacements for any of the spheres leads to the swimming distance $d$, and the sum of the forces $F_i$ correspond to the total external force $F$. Therefore,
\begin{align}
    T \dot \sigma = MF^2 + J(\epsilon,F)(\epsilon + Fd)= V(\epsilon,F)F + J(\epsilon,F) \epsilon,
\end{align}
as stated in the main text.

The local detailed balance between forward and backward rates requires that we account for the dissipation associated with each transition, so that
\begin{align}
    \frac{k_{\beta \alpha}}{k_{\alpha \beta}} = \exp\left[\frac{T \Delta \sigma_{\beta\alpha}}{k_B T}\right]=\exp\left[\frac{\Delta \epsilon_{\beta\alpha} + W_{\beta\alpha}}{k_B T}\right] =\frac{k_{0\beta \alpha}}{k_{0\alpha \beta}} \exp\left( \frac{W_{\beta \alpha}}{k_B T}\right),
\end{align}
where we have defined the work done by the external forces during the transition $\alpha\to\beta$ as $W_{\beta \alpha} \equiv \sum_i F_i \Delta x_{i,\beta\alpha}$. This gives the modification of the rates due to the external forces as stated in the main text.

\section{Results for negative chemical affinity}

In this section, we report results for negative values of cycle affinity $\epsilon$; see Figs. \ref{fig:currents_negative}, \ref{fig:entropy_negative}, and \ref{fig:tur_negative}. This implies, that the cycle will run in reverse in the absence of the force, and the swimmer will swim backwards (towards the left). The force $F$ still remains exerted on the first sphere (which now becomes the front of the swimmer). The sign convention for the force is still such that positive forces point towards the right, so positive forces $F$ in this case oppose the self-propulsion of the swimmer.
\begin{figure}[H]
\centering
\includegraphics[width=1.\linewidth]{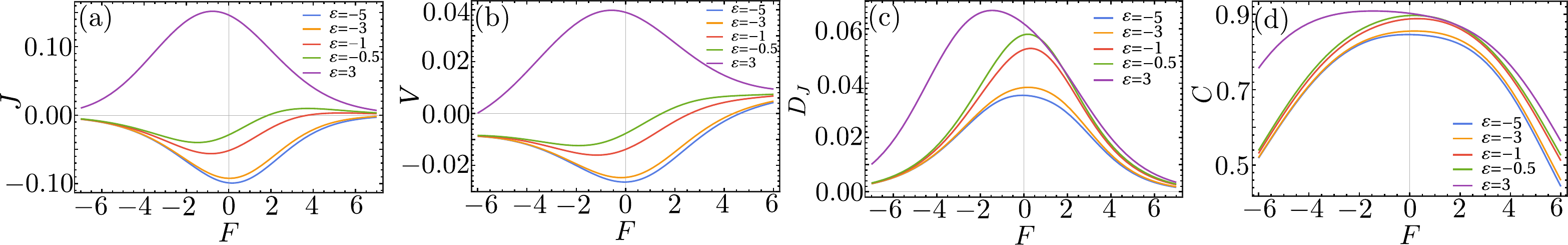}
\caption{Dependence on the external force $F$, for several values of the chemical affinity $\epsilon$, of (a) the chemical current $J$, (b) the spatial current or velocity $V$, (c) the chemical diffusion coefficient $D_J$, and (d) the correlation $C$ between $J$ and $V$.}
\label{fig:currents_negative}
\end{figure}
\begin{figure}[H]
\centering
\includegraphics[width=0.85\linewidth]{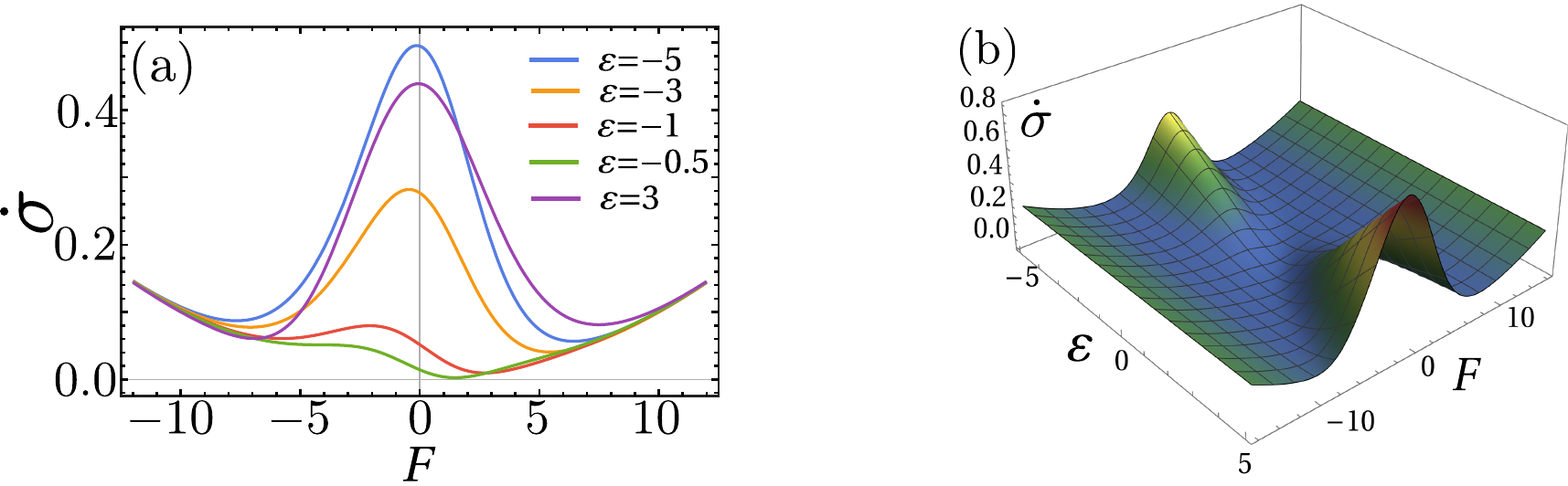}
\caption{(a) Entropy production rate $\dot{\sigma}$ as a function of the external force $F$, for several values of the chemical affinity $\epsilon$, and (b) as a function of both the external force and $\epsilon$.}
\label{fig:entropy_negative}
\end{figure}
\begin{figure}[H]
\centering
\includegraphics[width=0.95\linewidth]{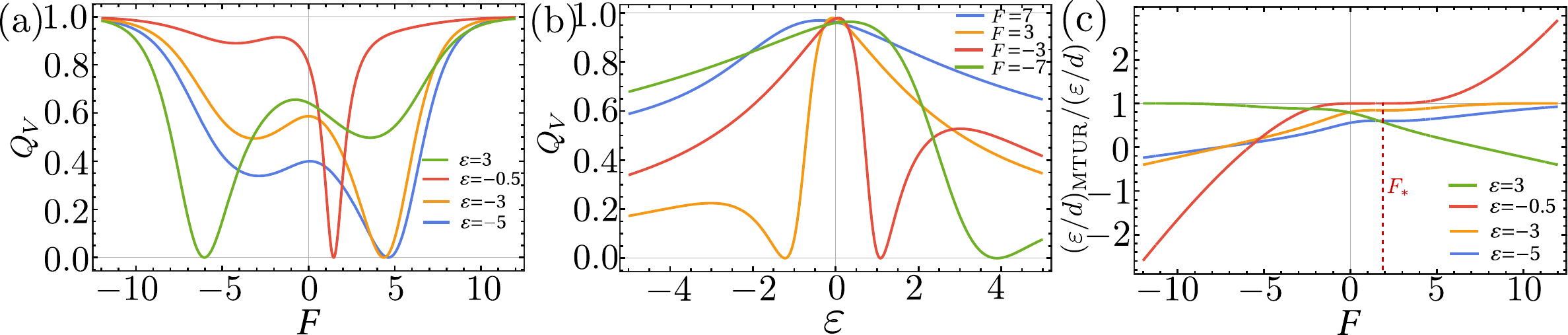}
\caption{(a,b) The quality factor $Q_V$ of the precision-dissipation tradeoff for the spatial current $V$, (a) as a function of $F$ for fixed values of $\epsilon$, (b) as a function of $\epsilon$ for fixed values of $F$. (c) Ratio of the MTUR-inferred [Eq.~(8) of the main text] and true chemical force $\epsilon/d$. $F_*$ is the critical force at which the chemical current vanishes. For $F<F_*$ the inferred value is a lower bound, for $F>F_*$ an upper bound, for $F=F_*$ it is exact.}
\label{fig:tur_negative}
\end{figure}
\end{widetext}
\end{document}